\documentclass{article}
\usepackage{spconf,amsmath,graphicx}

\title{teach an all-rounder with experts in different domains}
\name{Zhao You$^1$, Dan Su$^1$, Dong Yu$^2$}
\address{$^1$Tencent AI Lab, Shenzhen, China\\
$^2$Tencent AI Lab, Bellevue, WA, USA \\
\{dennisyou, dansu, dyu\}@tencent.com}
\begin{document}
%
\maketitle
\begin{abstract}
In many automatic speech recognition (ASR) tasks, an ideal model has to be applicable over multiple domains. In this paper, we propose to teach an all-rounder with experts in different domains. Concretely, we build a multi-domain acoustic model by applying the teacher-student training framework. First, for each domain, a teacher model (domain-dependent model) is trained by fine-tuning a multi-condition model with domain-specific subset. Then all these teacher models are used to teach one single student model simultaneously. We perform experiments on two predefined domain setups. One is domains with different speaking styles, the other is near-field, far-field and far-field with noise. Moreover, two types of models are examined: deep feedforward sequential memory network (DFSMN) and long short term memory (LSTM). Experimental results show that the model trained with this framework outperforms  not only multi-condition model but also domain-dependent model. Specially, our training method provides up to 10.4\% relative character error rate improvement over baseline model (multi-condition model).
\end{abstract}
\section{Introduction}
\label{sec:intro}

Thanks to deep learning approaches \cite{DE2, DE3}, great progress has been made in automatic speech recognition performance. Although deep neural networks have superior robustness over GMM systems on different conditions such as speaker, recording channel and acoustic environment \cite{DE4}, domain robustness is still a challenging problem. First, it is impractical to train a single model with good performance across all domains. Second, when encountering a new domain task, the models trained with other domain data are usually difficult to transfer knowledge to this new task.

Much effort has been devoted to solve these problems. One of the most effective and straightforward approach is multi-condition training or multi-style training.
Multi-condition training can trace back to \cite{MC4}, and has been shown to reduce mismatch for different noise conditions \cite{MC5}. Deep neural networks work particularly well with multi-condition training due to the large model capacity \cite{MC1, MC2, MC3}. Empirically, when sufficient amount of target domain data is available, performing fine-tuning on the multi-condition model with target domain data can produce excellent performance for the target domain specifically.

Recently, domain adaption based methods have been proposed for domain robustness.
Domain adaptation refers to the task of adapting models trained on one domain or mixed domains to the target domain. Previous study mainly focuses on the scenario where with limited target domain training data \cite{DA1}, one has to make a trade-off between test accuracy and the number of adaptation parameters. Further, domain adaptation for ASR is particularly difficult considering the mismatch in speaking styles,
noise types, and room acoustics etc \cite{DA2}.

Methods which augment DNN input with i-vector feature or speaker code have been developed against the speaker variations and channel variations. However, it is difficult for these methods to deal with other domain changes such as speaking style variations. For example, in spontaneous speech, the speaking rate is highly inconsistent and the articulation is highly variable, which are typically not observed in read speech. Moreover, the above approach has drawbacks which make it unsuitable for transferring knowledge across multiple domains\cite{DA3}.

In this paper, we propose a multi-domain teacher-student training framework for teaching an all-rounder with experts in different domains.
First, we train teacher models for each domain by fine-tuning a multi-condition model with the domain specific subset. Then, we exploit all these teacher models to train one single student model simultaneously. Consequently, the student model is an all-rounder across different domains. Our experimental results show that given sufficient amount of data for several domains the proposed method provides up to 10.4\% relative character error rate improvement over baseline models.


The rest of the article is organized as follows. The multi-domain teacher-student training method is described in Section 2. The experimental configuration is described in Section 3. We report the experimental results in Section 4 and conclude the paper in Section 5. 
 \section{multi-domain teacher-student training}
\label{sec:format}
\subsection{Teacher-student training}
In the teacher-student training framework, \cite{ts_ori1,ts_ori2,TS1} have shown that it is possible to train an student model to match the output distribution of a teacher model. 
Specially, the student model can be learned via single teacher network or multiple teacher networks. Details of the two learning methods will be discussed in the following section.
\subsubsection{single teacher network}
Teacher-student training is a general method for compressing acoustic model by minimizing the Kullback-Leibler divergence (KLD) between the output posterior distributions of the teacher model and the student model. That is, minimizing the loss function $\mathnormal{L_{KLD}}(\theta)$ defined as
\begin{equation}\label{equ:7}
 \mathnormal{L_{KLD}}(\theta) =-\sum_{l}\mathnormal{p_{t}(l|x)} log \mathnormal{p_{s}(l|x)}
 \end{equation}
where $ \mathnormal{p_{t}(l|x)} $ is the posterior probability of label $l$ given the input feature $x$ computed by the teacher model, and $\mathnormal{p_{s}(l|x)}$ is that computed by the student model. $\theta$ denotes the parameters of the student model. 

However, the valuable knowledge transferred from one teacher network to train the student network is limited. Thus, developing training approaches which use multiple teachers is needed.  

\subsubsection{multiple teacher networks}
In the case of multiple teacher networks, the performance of the student network is improved by leveraging information from multiple teachers. In \cite{multiple-ts2,multiple-ts3}, the outputs of multiple teacher networks are combined by weighted ensembles of posteriors from each teacher network as
\begin{equation}\label{equ:7}
\mathnormal{p_{t}(l|x)}= \sum_{k=1}^{N}\mathnormal{w_k}\mathnormal{p_{tk}(l|x)}
 \end{equation}
where $N$ is the number of teacher networks. $\mathnormal{p_{tk}(l|x)}$ is the posterior of the $k-th$ teacher network. $\mathnormal{w_k}$ is the interpolation weight. 

Though this approach allows the students to learn ensemble distribution created by multiple teachers, the characteristic of each teacher network is weakened by the interpolation process.
In that case, the student network can not obtain the most professional characteristic of each domain from the teacher network's knowledge. 

\subsection{Multi-domain training}
\vspace{-3pt}
\begin{figure}[!tb]
\begin{minipage}[b]{1.0\linewidth}
  \centering
  \centerline{\includegraphics[width=8.2cm]{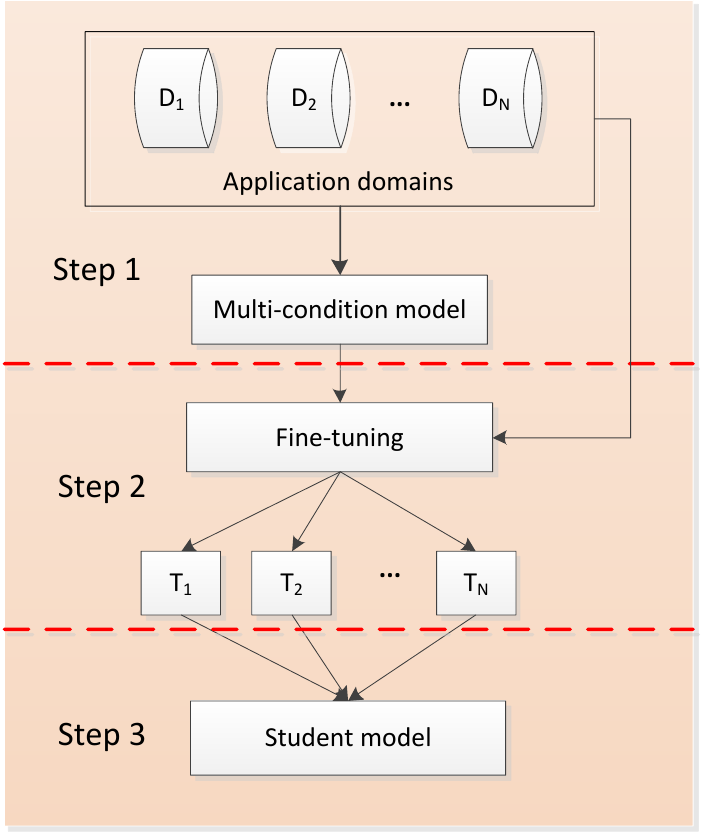}}
\end{minipage}

\caption{Training steps of multi-domain teacher-student learning.}
\end{figure}
To overcome the shortcomings of methods above, we propose a multi-domain teacher-student learning algorithm. In this method, each teacher network is trained with a domain specific data, which can be viewed as an expert of this domain and transfers the most professional characteristic of each domain.
The multi-domain teacher-student training framework is shown in figure 1. 
The process 
of training one student network from multiple teacher networks includes three steps:

1. We pool data from multiple application domains. $D_n$ denotes the $n$-th domain. Then, we train a multi-condition model with minibatches samples which are chosen randomly from the pooled set. 

2. Domain-dependent teacher models are produced by fine-tuning the multi-condition model with domain-dependent data respectively. $T_n$ denotes the $n$-th teacher model which is trained with the $n$-th domain data. 

3. The proposed model is learned from these $N$ domain-dependent teacher models. During the training process, samples in one minibatch are chosen randomly from the mixed data set, and may come from different domains. The train process exploits each sample for training by using the soft targets produced from its corresponding domain-dependent teacher model as in equation (3). Let $\mathnormal{p^{d}_{t}(l|x)}$ be the soft targets produced by domain specific data. $\mathnormal{\delta_{t}(l)}$ denotes the hard labels and $\mathnormal{w_{hard}}$ is its weight. Thus, $\mathnormal{p_{t}(l|x)}$ can be viewed as a linear interpolation of hard labels and soft labels. 

\begin{equation}\label{equ:7}
\mathnormal{p_{t}(l|x)}= (1-\mathnormal{w_{hard}})\mathnormal{p^{d}_{t}(l|x)} + \mathnormal{w_{hard}}\mathnormal{\delta_{t}(l)}
 \end{equation}

\label{ssec:subhead3}

\begin{table}
\caption{\textit{Total length (approx.hours) of each application domain for mixed speaking style corpus.} }
\label{tab:1}

\begin{center}
\begin{tabular}{|c|c|c|c|}
\hline
speaking-style & train &  dev & test  \\
\hline
Read       & 2k & 1 & 1   \\

Lect       & 2k & 1 & 1  \\

Spon       & 2k & 1 & 1 \\

 \hline
\end{tabular}
\end{center}
\end{table}

\begin{table}
\caption{\textit{Total length (approx.hours) of each application domain for mixed environment corpus. } }
\label{tab:1}

\begin{center}
\begin{tabular}{|c|c|c|c|}
\hline
environment & train &  dev & test  \\
\hline
Near          & 2k & 1 & 1   \\

Far           & 2k & 1 & 1  \\

FarNoise     & 2k & 1 & 1 \\

 \hline
\end{tabular}
\end{center}
\end{table}

\section{EXPERIMENTAL SETUP}
\label{sec:format}
\subsection{Training setup}
 The feature vectors used in all the experiments are 40-dimensional log-mel filterbank energy features appended with the first and second order derivatives. Log-mel filterbank energy features are computed with a 25ms window and shifted every 10ms. We stack 8 consecutive frames and subsample the input frames with 3. A global mean and variance normalization is applied for each frame. All the experiments are based on the CTC learning framework. We use the CI-syllable-based acoustic modeling method \cite{syllable} for the CTC learning. The target labels of CTC learning are defined to include 1394 Mandarin syllables, 39 English phones and a blank. Character error rate (CER) results are measured on the test sets. Decoding is performed with a beam search algorithm by using the weighted finite-state transducers (WFSTs).

\subsection{Datasets}
Our training corpus consists of a variety of application domains, all in Mandarin. In this work, we evaluate our approach on two kinds of domain setups. 
The number of utterances and the total length of each application domains are shown in Table 1 and Table 2. 

The first setup focuses on different speaking styles. We experiment on a mixed dataset with 3 kinds of different speaking styles, including read speech, lecture speech and spontaneous speech. We refer them as Read, Lect and Spon respectively. Each speaking style contains 2000 hours speech. The second setup focuses on the variation of environment, including near-field speech, simulated far-field speech and simulated far-field noisy speech. We refer them as Near, Far and FarNoise respectively. Together they contain a total of 6000 hours speech.  The far-field speech is generated using the image method described in \cite{far1}. A set of simulated room impulse responses (RIRs) are created with different rectangular room sizes, speaker positions and microphone positions, as proposed in \cite{far2}. Each environment contains 2000 hours speech. We hold out about 0.5\% (1 hours) as a development set for frame accuracy evaluation. Each test set includes 1k utterances which is about 1 hours of audio.

\subsection{Acoustic Model}
We present our work with DFSMN and LSTM acoustic models. The LSTM system uses 7 LSTM layers of 1024 cells, each with a recurrent projection layer of 512 units. For DFSMN model, we use 30 DFSMN components \cite{dfsmn}. 
The look-back order and lookahead order of each memory block is 5 and 1 respectively, and the strides are 2 and 1 respectively. 
For stable CTC learning, we clip gradients to [-1.0, 1.0].
We use the Kaldi \cite{kaldi} toolkit to train models and all models are trained in a distributed manner using BMUF \cite{BMUF1} optimization with 8 Tesla P40 GPUs.

\newcommand{\tabincell}[2]{\begin{tabular}{@{}#1@{}}#2\end{tabular}} 
\begin{table}
\caption{\textit{CER (\%) of 3 different training methods with DFSMN models. Results are with corpus of 3 different speaking styles.} }
\vspace{-3pt}
\label{tab:1}

\begin{center}
\begin{tabular}{|c|c|c|c|}
\hline
DFSMN & test-Read &  test-Spon & test-Lect  \\
\hline\hline
Baseline  &  17.37 & 23.18 &  15.92  \\
\hline
T1 (Read)   & 16.73 &  34.77&  26.77 \\

T2 (Spon)  & 26.95 &  21.98 &   21.57 \\

T3 (Lect) &  29.03 &  29.54 &  15.88  \\
\hline
student model & \tabincell{c} {\textbf{16.37} \\ \textbf{(-5.8\%)}} & \tabincell{c}{\textbf{20.76} \\ \textbf{(-10.4\%)}} &  \tabincell{c} {\textbf{15.13} \\ \textbf{(-5.0\%)}}  \\

\hline
\end{tabular}
\end{center}
\end{table}

\section{EXPERIMENTAL RESULTS}
In this work, we evaluate the performance of the proposed method on several large vocabulary Mandarin speech recognition tasks including near-field speech and far-field speech as described in section 3.2.
\label{sec:format}
\subsection{Mixed speaking style corpus}
For the first set of experiments, we validate the effectiveness of the proposed method by dealing with mixed speaking style speech. To find an appropriate value for $\mathnormal{w_{hard}}$ in equation (3), we randomly select about 25 \% data from each domain to constitute a training set and perform the experiment. We
find that $\mathnormal{w_{hard}}=0.8$ achieves best performance on the development set. Thus, we set $\mathnormal{w_{hard}}=0.8$ for all the experiments.

Table 3 shows the performance comparison on mixed speaking style corpus with DFSMN acoustic models. Line 2 presents the results of the baseline system (multi-condition model). The following three lines present results of 3 teacher models trained on $Read$, $Lect$ and $Spon$ data respectively. The last line presents results of the student model which is learned by using 3 teacher models. Compared with the baseline system, the teacher models work well when the test domains match the model domains while poorly when the domains mismatch. Specially, the results clearly show that the student model performs best compared with other domain-adaptation (teacher) models. Finally, our proposed method achieves up to 10.4\% relative CER improvement over the baseline model on the test-Spon test set. Table 4 shows the corresponding results of LSTM networks. It can be concluded that the student network still outperforms the baseline system on all the domain tests. However, the gain is smaller compared with DFSMN models. 
\begin{table}
\caption{\textit{CER (\%) of 3 different training methods with LSTM models. Results are with corpus of 3 different speaking styles.} }
\label{tab:1}

\begin{center}
\begin{tabular}{|c|c|c|c|}
\hline
LSTM & test-Read &  test-Spon & test-Lect  \\
\hline\hline
Baseline    & 17.49 & 21.09 & 15.49   \\
\hline
T1 (Read)   & 17.23 & 29.78 & 22.11  \\

T2 (Spon)    & 23.63 & 19.68 & 19.75   \\

T3 (Lect)   & 25.25 & 25.92 & 14.58   \\
\hline
student model   & \tabincell{c} {\textbf{16.79} \\ \textbf{(-4.0\%)}} & \tabincell{c}{\textbf{19.85}\\ \textbf{(-5.9\%)}} & \tabincell{c} {\textbf{14.99} \\ \textbf{(-3.2\%)}}   \\

\hline
\end{tabular}
\end{center}
\end{table}

\begin{figure}[!tb]
\begin{minipage}[b]{1.0\linewidth}
  \centering
  \centerline{\includegraphics[width=8.0cm]{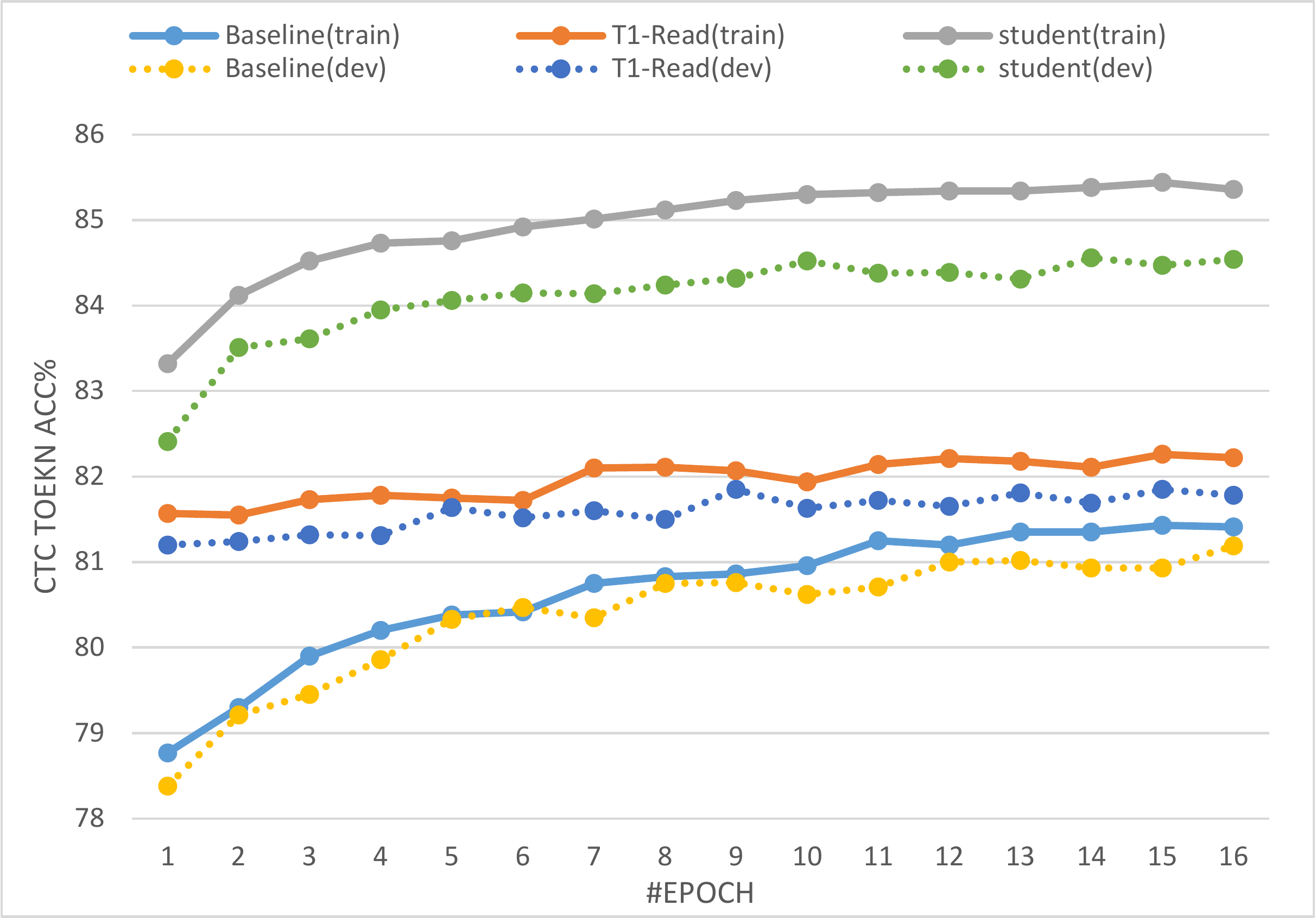}}
\end{minipage}

\caption{Frame accuracy on $Read$ train and development set against the number of epochs through the training dataset for 3 different training methods.}
\end{figure}
\vspace{-3pt}



Figure 2 shows the frame accuracy of different models. As shown, the "student" model produces the highest frame accuracy compared with both the baseline model and the "teacher" model. Since the "student" model is trained on the linear interpolation of the hard targets and soft targets, this suggests that the soft targets role as a significant contribution to train the student model of high accuracy.

\begin{table}
\caption{\textit{CER (\%) of 3 different training methods with DFSMN models. Results are with corpus of 3 different far-field environments.} }
\label{tab:1}

\begin{center}
\begin{tabular}{|c|c|c|c|}
\hline
DFSMN & test-Near &  test-Far & test-FarNoise  \\
\hline\hline
Baseline            & 17.08 & 26.04 & 46.28   \\
\hline
student model  & \tabincell{c} {\textbf{15.76}\\ \textbf{(-7.7\%)}} & \tabincell{c}{\textbf{23.44} \\ \textbf{(-10.0\%)}} & \tabincell{c} {\textbf{42.47} \\ \textbf{(-8.2\%)}}   \\
\hline
\end{tabular}
\end{center}
\end{table}

\begin{table}
\caption{\textit{CER (\%) of 3 different training methods with LSTM models. Results are with corpus of 3 different far-field environments.} }
\label{tab:1}

\begin{center}
\begin{tabular}{|c|c|c|c|}
\hline
LSTM & test-Near &  test-Far & test-FarNoise  \\
\hline\hline
Baseline            & 17.04 & 26.36 & 44.03   \\
\hline
student model  &\tabincell{c} {\textbf{16.74} \\ \textbf{(-1.8\%)} } & \tabincell{c} {\textbf{25.60} \\ \textbf{(-2.9\%)}} & \tabincell{c} {\textbf{43.00} \\ \textbf{(-2.3\%)}}   \\
\hline
\end{tabular}
\end{center}
\end{table}


\vspace{-10pt}
\subsection{Mixed near-field and far-field corpus}
To comprehensively validate the effectiveness of our proposed method, we also investigate the performance on the mixed near-field and far-field corpus. The multi-condition model, student model and teacher models have the same configuration with that used in Part 4.1.
Table 5 shows that the DFSMN student model trained by our proposed method also significantly outperforms the baseline system. In particular, our training method provides up to 10\% relative CER improvement over the baseline model on the test-Far test set. 
This shows that our proposed method can achieve consistent improvements, no matter on a mixed speaking styles corpus or a mixed near-field and far-field corpus. 
Table 6 shows the corresponding results of LSTM student model. An observation is that the improvement on LSTM models is smaller compared with DFSMN models. 

\vspace{-10pt}
\section{CONCLUSIONS AND FUTURE WORKS}
\label{sec:pagestyle}

In this paper, we propose 
a multi-domain teacher-student training method for teaching an all-rounder with experts in different domains. 
We explore this method for acoustic modeling on two different tasks. We find that the model trained by this method not only outperforms multi-condition model but also outperforms the domain-dependent model produced by fine-tuning a multi-condition model with the target domain data set. Table 3 shows that our method achieves up to 10.4\% relative CER improvement over the baseline model. 

\cite{multiple-ts3} has shown that combining intermediate representations of multiple teacher networks can significantly improve the student network's performance. Thus, we will explore this training strategy to improve the performance of LSTM models in the future work. 




\vfill\pagebreak

%

\bibliographystyle{IEEEbib}
\bibliography{strings,refs}

\end{document}